# Plane photoacoustic wave generation in liquid water using irradiation of terahertz pulses

Masaaki Tsubouchi,[1,*] Hiromichi Hoshina,[2] Masaya Nagai,[3] and Goro Isoyama[4]

[1]Kansai Photon Science Institute, National Institutes for Quantum and Radiological Science and Technology (QST), 8-1-7 Umimedai, Kizugawa, Kyoto 619-0215, Japan
[2]RIKEN Center for Advanced Photonics, RIKEN, 519-1399 Aramaki-Aoba, Aoba-ku, Sendai, Miyagi 980-0845, Japan
[3]Graduate School of Engineering Science, Osaka University, 1-3 Machikaneyama, Toyonaka, Osaka 560-8531, Japan
[4]Institute of Scientific and Industrial Research (ISIR), Osaka University, 8-1 Mihogaoka, Ibaraki, Osaka, 567-0047, Japan

*Author to whom correspondence should be addressed: *tsubouchi.masaaki@qst.go.jp*

**ABSTRACT**
We demonstrate photoacoustic wave propagation with a plane wavefront in liquid water using a terahertz (THz) laser pulse. The THz light can effectively generate the photoacoustic wave in water because of strong absorption via a stretching vibration mode of the hydrogen bonding network. The excitation of a large-area water surface irradiated by loosely focused THz light produces a plane photoacoustic wave. This is in contrast with conventional methods using absorbers or plasma generation using near-infrared laser light. The photoacoustic wave generation and plane wave propagation are observed using a system with a THz free-electron laser and shadowgraph imaging. The plane photoacoustic wave is generated by incident THz light with a small radiant exposure of <1 mJ/cm$^2$ and delivered 600 times deeper than the penetration depth of THz light for water. The THz-light-induced plane photoacoustic wave offers great advantages to non-invasive operations for industrial and biological applications as demonstrated in our previous report (Yamazaki et al, Sci. Rep. 846295 (2020)).

Pressure wave generation is one of the important processes induced by laser irradiation in liquid media.[1-4] The photoacoustic wave with the sound velocity has been investigated in water for non-invasive tomographic imaging for biomedical issues in water.[5-8] The shockwave, which is a hypersonic wave with high pressure, has also been examined for medical applications, such as drug delivery and microscale cell manipulation.[3,9-14] When laser light with short pulse duration is strongly absorbed by liquid or chromophores dissolved in liquid, the energy of the light is instantaneously confined in a small volume. Subsequently, the energy is released as a pressure wave into the liquid via the thermoelastic effect, when the stress confinement condition, $\tau \alpha v_s \ll 1$, is satisfied, where $\tau$ is the laser pulse duration, $\alpha$ is the absorption coefficient of liquid, and $v_s$ is the speed of sound in liquid.[9] However, the widely used types of pulsed laser light are at visible or near-infrared (IR) wavelengths to which water is transparent ($\alpha < 0.1$ cm$^{-1}$). Therefore, an absorber, *e.g.*, black rubber, fluorescent dye materials, and so on, is required. By contrast, mid- and far-IR light is strongly absorbed by water with $\alpha > 100$ cm$^{-1}$ at wavelengths of 3–100 μm. A carbon-dioxide laser with a wavelength of 10.6 μm can provide strong mid-IR light pulses, but cannot satisfy the stress confinement condition because of the long pulse duration $\tau > 10$ ns. Even if the stress confinement condition is not satisfied, tightly focused laser pulses can generate a shockwave following plasma generation in water.[15-17] These processes are summarized in Fig. 1(A).

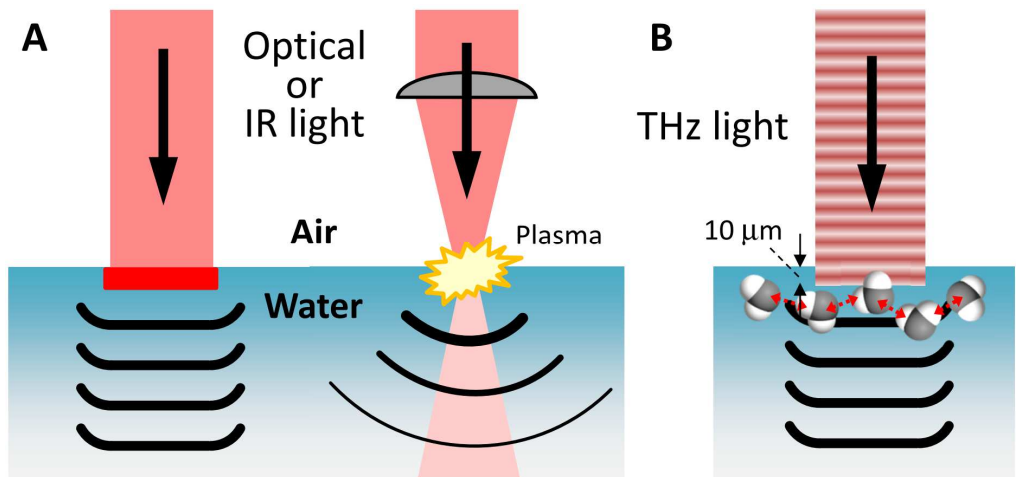

**Fig. 1.** Mechanisms of pressure wave generation at the air–water interface. (**A**) An optical or IR laser generates pressure waves via either by a light absorber or plasma generation by strong laser light. (**B**) A THz laser, in contrast, can directly generate a plane wave from a relatively weak field with a loose focus.

In previous methods of pressure wave generation, it was necessary to introduce an absorber or to produce the plasma in water or tissues, which are invasive processes with serious risk of damage to materials and biological tissues. To overcome this problem, we propose terahertz (THz) light-induced photoacoustic wave generation as shown in Fig. 1(B). The THz light has a frequency of $10^{12}$ Hz, which lies between the frequency ranges of light and radio waves. Due to the resonance of the intermolecular vibration in the hydrogen-bonding network of liquid water around 5 THz,[18,19] the THz light is completely absorbed very close to the surface of the water, with a penetration depth of 10 μm. The strong absorption of the THz laser light with a picosecond pulse duration induces a rapid and local pressure increase followed by effective photoacoustic wave generation without requiring any additional absorber. The strong absorption of the THz light also achieves plane wave propagation. Because according to Huygens' principle a large-area excitation source is required for plane wave propagation, efficient energy conversion from the light to the pressure wave due to the strong absorption is necessary. A plane wave is superior to a spherical one for long-distance propagation without intensity drop and for reconstruction of tomographic images. In addition, the low photon energy (20 meV at 5 THz) of the THz light does not induce any ionization, dissociation, or structural changes in molecules. THz light thus has great advantages for non-destructive pressure-wave generation in industrial and medical applications. The plane acoustic wave generation other than the laser induced method has been also achieved by the piezo transducer and applied to manipulate the cultured cell in water.[20] This method can also prevent damage to tissues. But the transducer has to be contacted to the water surface or embedded in water, which is not stable for the medical applications.

Recently, we observed the demolition of actin filaments in water and living cells irradiated by the THz light.[21] We presumed that the demolition was the result of the pressure wave propagation in water induced by the THz light. In this study, we clarify photoacoustic wave generation with picosecond THz light pulses provided by a free-electron laser (FEL) and detect it using the shadowgraph imaging method with 10 ns temporal resolution and 10 μm spatial resolution. The characteristics of the THz photoacoustic wave are investigated by observing the spatiotemporal evolution.

**Methods**
**THz free-electron laser.** For the THz light source to generate the photoacoustic wave, we employed the THz-FEL on the L-band electron linear accelerator (LINAC) at the Research Laboratory for Quantum Beam Science, Institute of Science and Industrial Research, Osaka University.[22-26] The characteristics and the evaluation method for THz pulses from the FEL were described in detail in the previous paper.[23,26] Linearly polarized THz macropulses are generated by the THz-FEL at a repetition rate of 5 Hz with the highest pulse energy of 50 mJ. Figure 2(A) shows a THz macropulse structure measured with a fast pyroelectric detector (Molectron P5-00). The peak intensity of each micropulse is normalized by the summation of the peak intensities of all micropulses in the macropulse. The macropulse contains a train of approximately 220 micropulses separated at 36.9 ns intervals (27 MHz repetition). The highest micropulse energy was estimated to be 340 μJ, which is by far the largest THz-FEL micropulse energy reported in the world. The temporal width of the micropulse was measured to be 1.7 ps by an electro-optic sampling technique (see the supplementary information and Fig. S2).[27,28] The center frequency is tunable in the range of 3–7 THz, which corresponds to a lower frequency edge of the absorption band due to the intermolecular vibration in liquid water (see Supplementary Fig. S1). Within the bandwidth of the THz light (~ 0.6 THz), the absorbance is not significantly changed. At 4 THz, the absorption coefficient of liquid water is 800 cm$^{-1}$,[18] which implies that more than 99.7% of irradiated energy is absorbed within 0.1 mm of the surface.

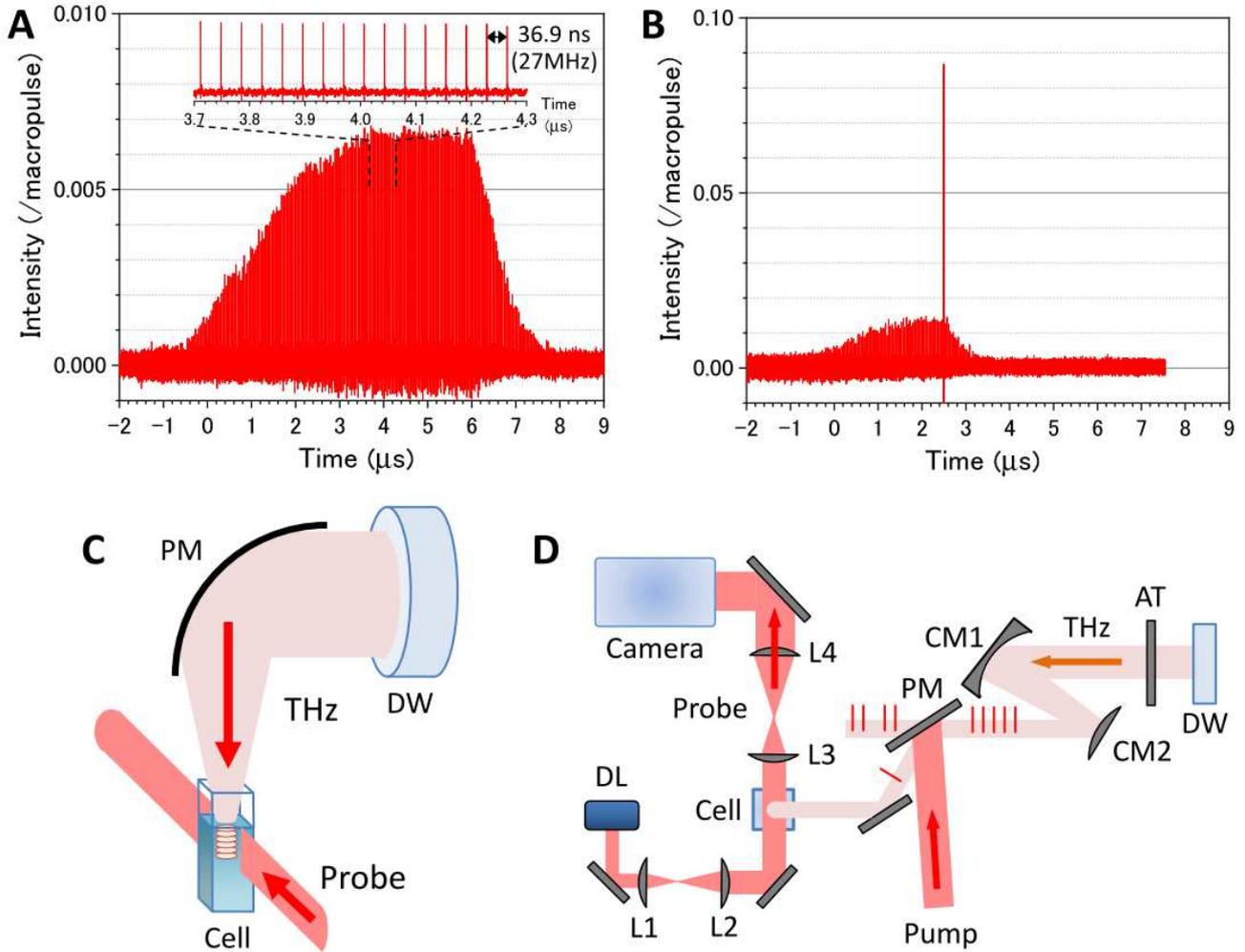

**Fig. 2.** Generation and observation schemes for THz-light-induced photoacoustic wave propagation in water. (**A**) Macropulse structure containing a train of approximately 220 micropulses. Inset shows the enlarged micropulse train with an interval of 36.9 ns (a repetition rate of 27 MHz). (**B**) Result of a single micropulse pick-up by the GaAs plasma mirror with a nanosecond gating. In (A) and (B), the peak intensity of each micropulse is normalized by the summation of the peak intensities of all micropulses in the macropulse. (**C**) Experimental setup for generating and probing the THz-FEL-induced photoacoustic wave on an air–water interface. DW: polycrystalline diamond window; PM: off-axis gold-coated parabolic mirror with a focal length of 50 mm; and Cell: quartz sample cell on a lab jack to adjust the focus diameter of the THz light on the air–water interface. (**D**) Optical layout for single THz-pulse pick-up and shadowgraph imaging systems. DL: diode laser; Camera: image-intensified CCD camera; AT: THz attenuator; PM: plasma mirror with nanosecond gating; L1, L2, L3, and L4: BK7 non-coated spherical mirrors with focal lengths of 50, 150, 100, and 200 mm, respectively; CM1 and CM2: gold-coated concave and convex mirrors with focal lengths of 200 and −100 mm, respectively.

**Single THz-pulse pick-up.** Figure 2(B) shows the single THz-micropulse picked up from the pulse train by the plasma mirror with nanosecond gating.[29,30] We employed a GaAs wafer irradiated by an intense femtosecond Ti:sapphire laser pulse as the nanosecond plasma mirror. The GaAs is transparent to THz light and has a Brewster angle of 75°. When the GaAs wafer is irradiated by the near-IR light with the energy of 500 μJ and the spot size of 1 cm$^2$, an electron plasma with the density of 10$^{19}$ cm$^{-3}$ is generated on the surface with a minority carrier lifetime of less than 10 ns.[31,32] Then, the GaAs plasma mirror can pick up only a single micropulse from the pulse train with a time interval of 36.9 ns. Since residual pulses remained as a result of the small difference from the Brewster angle, the electron beam injection was stopped just after the pick-up pulse. As shown in Fig. 2(B), the energy ratio of the picked-up single pulse to the macropulse was 8.7%. This is 13 times larger than that of micropulses in the original macropulse shown in Fig. 2(A). The detail of the single THz-pulse pick-up is described in the Supplemental information.

**Shadowgraph imaging system.** Figures 2(C) and (D) show a schematic diagram of the photoacoustic wave generation and observation system. The THz light passed through a polycrystalline diamond window from the evacuated FEL system into the air. A single THz micropulse was picked up by the GaAs wafer pumped by Ti:sapphire laser light (800 nm wavelength, 0.5 mJ pulse energy, 100 fs pulse duration, and 10 mm diameter) synchronized to the timing of the FEL macropulse generation, and loosely focused on the distilled water sample using a gold-coated off-axis parabolic mirror with a 50 mm focal length. To evaluate the spot size of the THz pulse on the water surface, we used the knife-edge method. The input pulse energy was attenuated with THz attenuators (TYDEX), which contained wedged silicon wafers with different attenuation levels.

A two-dimensional cross-section image of the photoacoustic wave was observed using the shadowgraph technique, which clearly shows an inhomogeneous density distribution in transparent media.[33] In the shadowgraph image, the signal intensity depends on the second derivative of the refractive index, which is related to pressure and density via the Gladstone–Dale relation, $\rho \propto n - 1$, where $\rho$ is the density and $n$ is the refractive index. Therefore, the shadowgraph is sensitive to the pressure wave, that is, the photoacoustic wave. As a probe light, a CW diode laser (LDM670, Thorlabs) with an output wavelength of 670 nm irradiates the distilled water in the quartz sample cell with a thickness of 10 mm. The probe light was incident on the water sample perpendicular to the photoacoustic wave propagation and was imaged by a 4$f$-type lens system onto the image-intensified CCD of a Princeton PI-MAX3 camera. The image capturing system was synchronized to the FEL and gated with a time duration of 10 ns. The time gate was electronically scanned with the delay generator in the PI-MAX3 system. In this system, we observed time evolution of the THz-light-induced phenomena from a nanosecond to a millisecond time scale with a time resolution of 10 ns. The background image obtained without the THz light irradiation was subtracted from the original images and the resultant background-subtracted images are shown in the figures.

## Results and Discussion

**Measurement of THz-light-induced photoacoustic waves.** Figure 3(A) shows a shadowgraph image of a water sample irradiated by the THz-FEL with a macropulse energy of 2.6 mJ at the center frequency of 4 THz. This energy corresponds to an average micropulse energy of 18 μJ, that is, a radiant exposure of 4.6 mJ/cm$^2$ with a beam diameter of 0.7 mm at the water surface. The radiant exposure is defined as the single micropulse energy per unit area. A stripe pattern is clearly seen in the image. Each horizontal line corresponds to a pulse front of a photoacoustic wave induced by the THz pulse train shown in Fig. 2(A). A neighboring line is a photoacoustic wave generated by an adjacent THz pulse. Thus, the propagation of the photoacoustic wave in water can be obtained from a single captured image. One remarkable feature is that the THz-light-induced photoacoustic wave has a plane wavefront. The plane photoacoustic wave is generated from the plane source with loosely focused THz-FEL light, because its beam diameter of 0.7 mm is considerably larger than the thickness of the photoacoustic wavepacket, < 10 μm, estimated from the width of the each horizontal line in Fig. 3(A). The nature of the plane wave causes the long-distance propagation of the photoacoustic wave, as explained in Fig. 1(B). The spacing between wavepackets is 55 μm on average, which corresponds to the distance travelled by the photoacoustic wave in the time intervals of the THz pulse train, 36.9 ns. Thus, we can estimate the speed of the photoacoustic wave in water to be 1491 m/s, which is the same as the sound velocity in distilled water at 23 °C.

Figure 3(B) displays the trains of photoacoustic waves measured by scanning the gate timing of the CCD camera. The amplitude is obtained from the horizontal sum of the pixel intensities in each row of the shadowgraph image. The original sequential images are shown as a movie in Supplementary Video S1 online.

The photoacoustic waves arise at the air–water interface with time intervals of 36.9 ns and propagate deeper into the water.

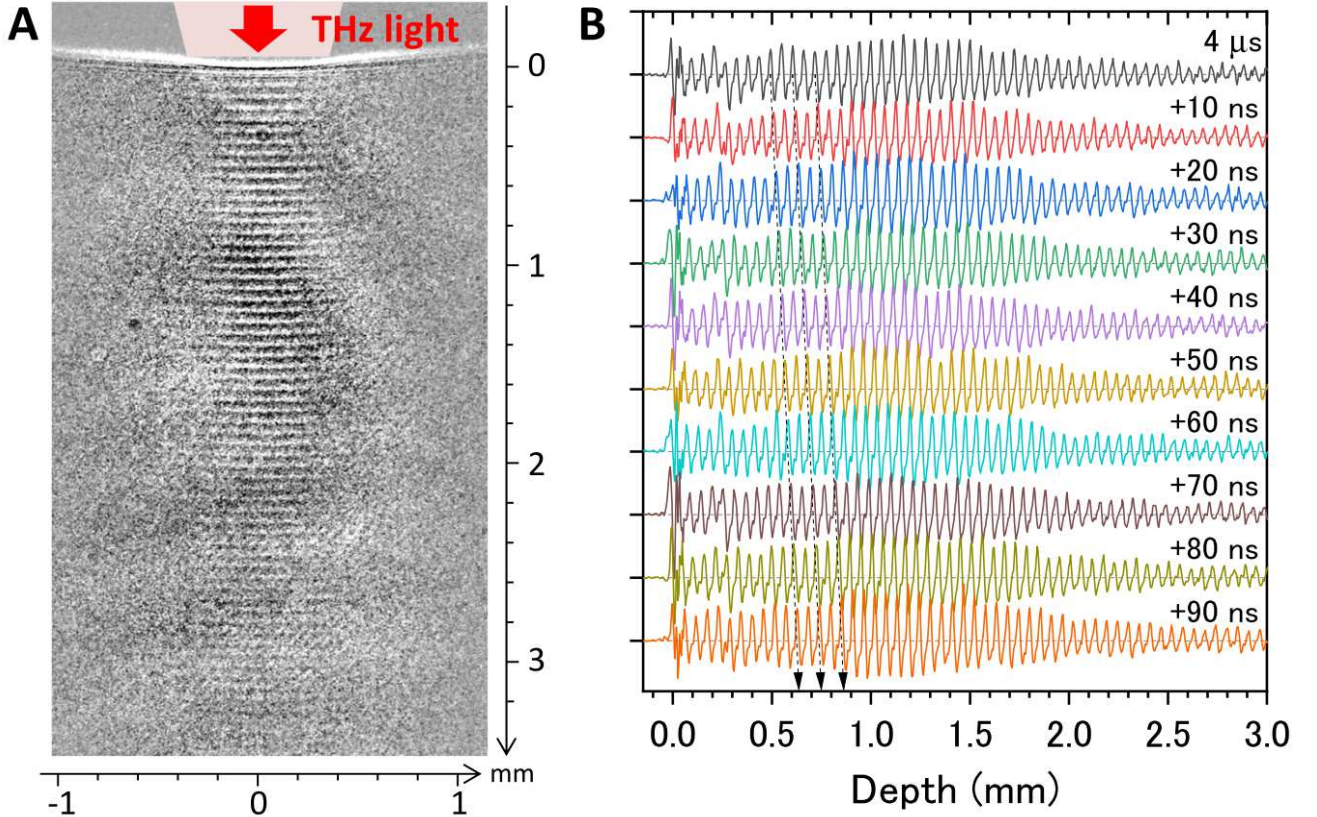

**Fig. 3.** Photoacoustic wave propagation induced by the THz pulse train. (**A**) Snapshot image of a train of photoacoustic waves induced by the THz-FEL with a frequency of 4 THz and an average micropulse energy of 18 μJ. This image was taken with a time gate of 10 ns. A movie consisting of sequential snapshot images is shown in Supplementary Video S1 online. (**B**) Spatiotemporal propagation of photoacoustic waves. The wave amplitudes are shown as a function of depth. The amplitude is obtained from the horizontal sum of the pixel intensities in each depth of the image. The dashed arrows trace propagation of specific three photoacoustic waves.

**Characteristics of photoacoustic wave propagation in water.** To quantitatively analyze the generation and propagation processes of the photoacoustic wave, the waveforms shown in Fig. 3(B) as a function of depth, $z$, were Fourier transformed to obtain the amplitudes as a function of wavenumber, $k$, which is related to speed as follows: $v_s = (k\Delta t)^{-1}$, where $\Delta t$ is the time interval of the micropulses. Figure 4(A) summarizes the speed of the photoacoustic wave as a function of depth for different THz frequencies and THz micropulse energies. The frequency and energy dependencies could not be clearly seen in the results, even though the absorption coefficient varies from 600 cm$^{-1}$ at 3.2 THz to 1150 cm$^{-1}$ at 5 THz and the radiant exposure varies from 1 to 10 mJ/cm$^2$. According to the weak shock theory, the speed of the pressure wave, $v_p$, depends on the local pressure increase $p$ as $v_p = v_s + \beta p/2\rho v_s$, where $\beta$ and $\rho$ are the nonlinearity and the density of water, respectively.[4,34] When the pressure wave is generated by the thermoelastic effect from the photoexcited region, the local pressure increase is expressed to be $p(z = 0) = \Gamma\alpha F$, where $F$ is the incident radiant exposure, $\Gamma = \gamma v_s^2/c_p$ is the Grüneisen coefficient of water, $\gamma$ is the coefficient of volume expansion, and $c_p$ is the heat capacity at constant pressure.[9] In our experimental conditions, the local pressure increase is estimated to be 0.5 MPa at $\alpha = 800$ cm$^{-1}$ and $F = 5$ mJ/cm$^2$, and thus the increase of wave speed, $v_p - v_s = \beta p/2\rho v_s$, is only 0.6 m/s. This is much smaller than our experimental accuracy, which supports our result.

The above discussion suggests that the local pressure, that is, the amplitude of the photoacoustic wave, is proportional to the product of the absorption coefficient and the radiant exposure of the THz light. To verify this expectation, the Fourier amplitude of the photoacoustic wave is plotted as a function of $\alpha F$ in Fig. 4(B). The result is well reproduced by the model based on the thermoelastic effect. Another significant feature is that the speed of the photoacoustic wave seems to decrease with the propagation distance. The wave speeds are $v_s = $ 1518 m/s ($z = 0.5$ mm) and 1492 m/s ($z = 3$ mm), which correspond to the speed of sound at temperatures of 34 °C and 23 °C, respectively.[35] One of the reasons for the deceleration of the photoacoustic wave in water is the spatial gradient of the temperature. However, the macropulse structure consisting of 220 micropulses complicates quantitative analysis of the photoacoustic wave generation and propagation due to an excessive supply of light energy to the water surface, interference between adjacent photoacoustic waves, and so on. Therefore, we tried to pick up a single acoustic wave from the wave train.

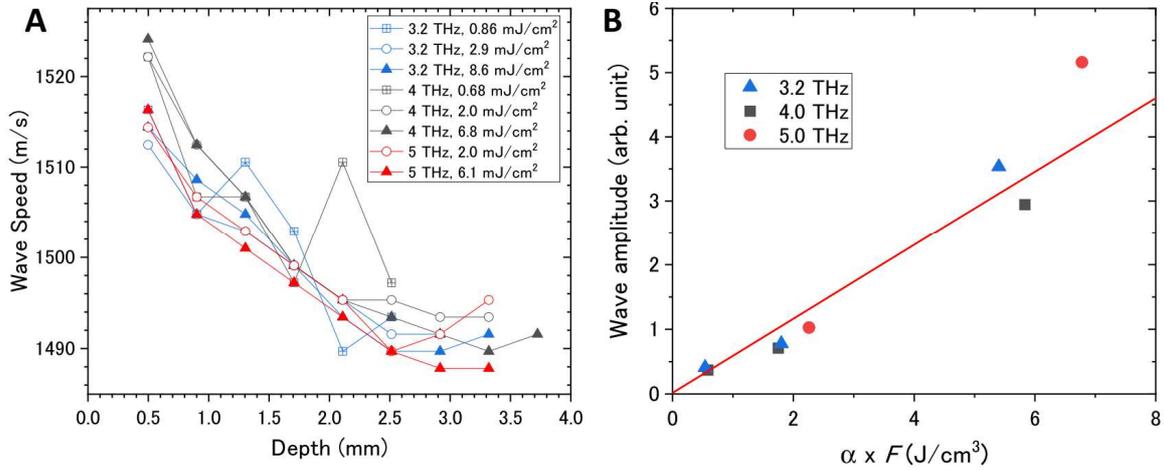

**Fig. 4.** Quantitative analysis of photoacoustic waves. (**A**) Speed of the photoacoustic waves as functions of depth for different THz frequencies and THz micropulse energies. The Fourier transform was calculated at a depth of $z \pm 0.5$ mm. (**B**) Amplitude of the photoacoustic wave as a function of the product of the absorption coefficient, $\alpha$, and the radiant exposure, $F$, of the THz light. The red line is the fitting of $p \propto \alpha F$.

**Single THz-pulse pick-up.** Figures 5(A) to (D) show shadowgraph images of the photoacoustic wave induced by the single THz pulse shown in Fig. 2(B). The center frequency and the estimated single pulse energy are 5 THz and 110 µJ, respectively. The radiant exposure is calculated to be 90 mJ/cm$^2$ with a beam diameter of 0.4 mm. A movie consisting of sequential snapshot images is shown in Supplementary Video S2 online. As mentioned before, the energy ratio of the residual THz pulses to the single pulse decreases to 1/13 of the original macropulse by the plasma mirror pick-up method. Therefore, the thermal effect due to the residual pulses can be suppressed significantly. The propagation of the single photoacoustic wave is clearly seen in the images. We emphasize that the photoacoustic wave reaches 6 mm in depth with an angular spread of 3°, which is 600 times longer than the skin depth of water for THz light. This result indicates that the energy of the THz light is delivered into the deep water by the photoacoustic wave as mechanical energy. Huygens' principle suggests that the plane wave is propagated when the diameter of the generating area is enough larger than the wavelength. The THz light induced photoacoustic wave is not the continuous wave but the pulse propagating with the sound velocity. Then, the wavelength of the photoacoustic wave is considered to be the order of the packet size, ~ 10 µm. The spot size of the THz light focused on the water surface is 0.4 mm which is much larger than the packet size, that is, the wavelength. Therefore, even when we use the pick-up single THz pulse is used for the photoacoustic wave generation, the plane wave can be generated as seen in Fig. 5.

The wave speed can be obtained from the time propagation of the wave in the series of images, and is summarized in Fig. 5(E). Any depth and frequency dependencies cannot be seen in the result, in contrast to that induced by the THz pulse train. Therefore, the deceleration of the photoacoustic wave induced by the THz pulse train might be due to a temperature increase close to the surface by injection of excess energy. By contrast, the wave speed of 1496 m/s is not significantly changed from that obtained in the THz pulse train, although the incident radiant exposure of the single THz pulse is twenty times larger than that per pulse in the THz pulse train. This fact also supports the former discussion of the equation, $v_p - v_s = \beta p / 2\rho v_s$.

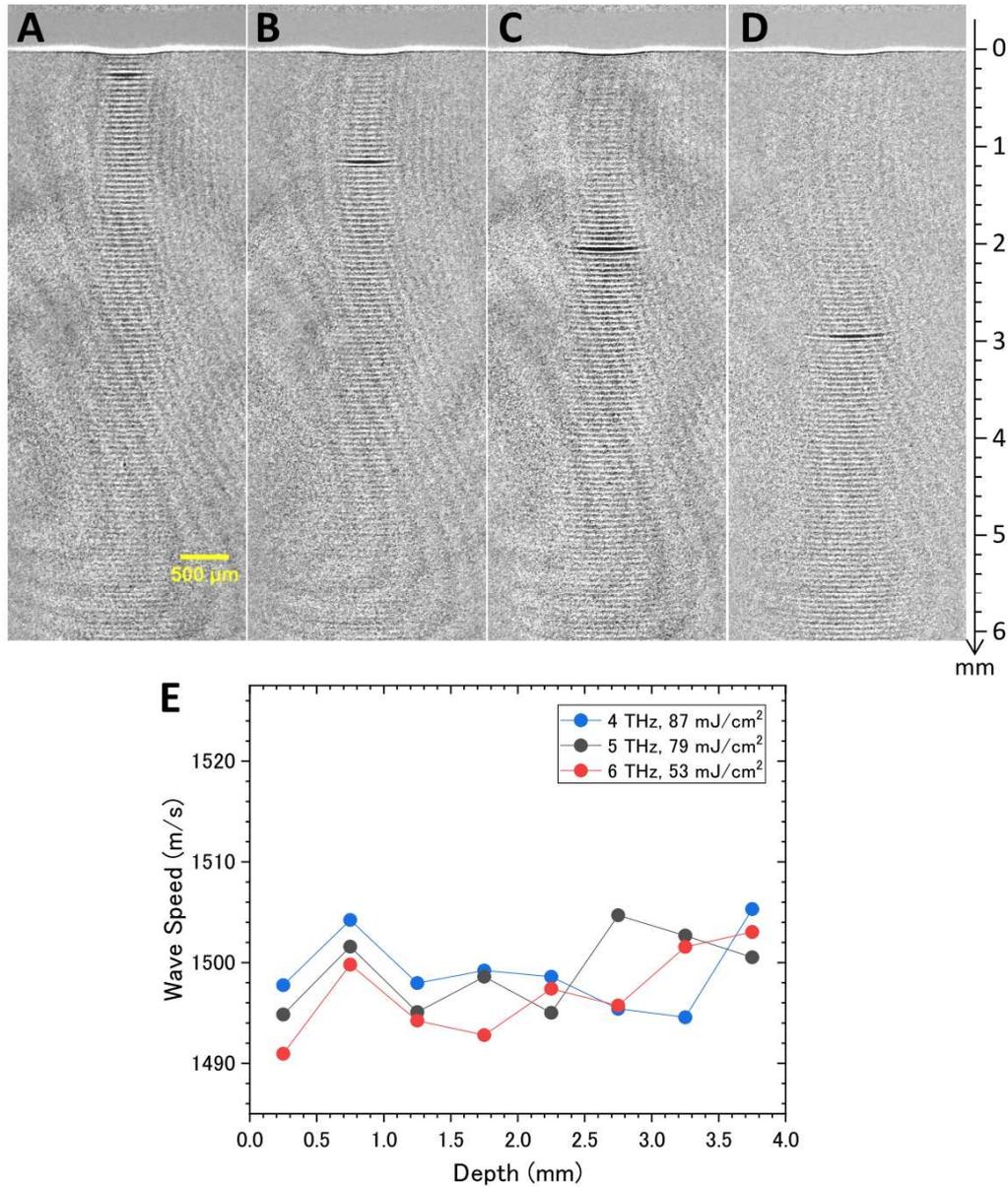

**Fig. 5.** Photoacoustic wave propagation induced by a single THz pulse. Snapshot images of (**A**) the photoacoustic wave induced by the THz-FEL with a frequency of 5 THz and at delay times of (**B**) 60, (**C**) 120, and (**D**) 180 ns. These images were taken with a time gate of 10 ns. A movie consisting of sequential snapshot images is shown in Supplementary Video S2 online. (**E**) Speed of the photoacoustic waves as functions of depth for different THz frequencies.

**Cavitation bubble formation triggered by photoacoustic wave generation.** Finally, we discuss THz-FEL-induced phenomena after the photoacoustic wave leaves from the water surface. Figure 6 shows a series of shadowgraph images captured from microsecond to millisecond time scales. By using the THz pulse train, a bubble core appears at the surface after the train of photoacoustic waves passes through the viewing area and grows into a hemispherical shape with a submillimeter diameter over several hundred microseconds. After 1 ms, the bubble collapses as a result of surface tension. This cavitation bubble formation has been also studied in shockwave generation by near- and mid-IR laser light.[4,9] There have been several proposals to employ laser-induced cavitation bubbles for cell surgery,[3,36] colloid processing,[37] and so on. Figures 6(E) to (H) show a series of shadowgraph images of water excited by a single THz pulse. The air–water interface is pushed down as the photoacoustic wave is generated, and a depression of 50 μm persists after THz light irradiation. The depression and the cavitation bubble generation in the air–water interface after laser-induced pressure wave generation has been discussed in previous studies.[3,9,36] As the pressure wave propagates into the water from the surface, water is excluded from the surface by the pressure wave. Then, the pressure at the interface is rapidly reduced, which enhances vaporization due to the lowering of the boiling point in addition to rapid heating of the water. However, a single THz pulse cannot form a bubble core at the air–water interface, which may be due to the lack of total incident THz light energy.

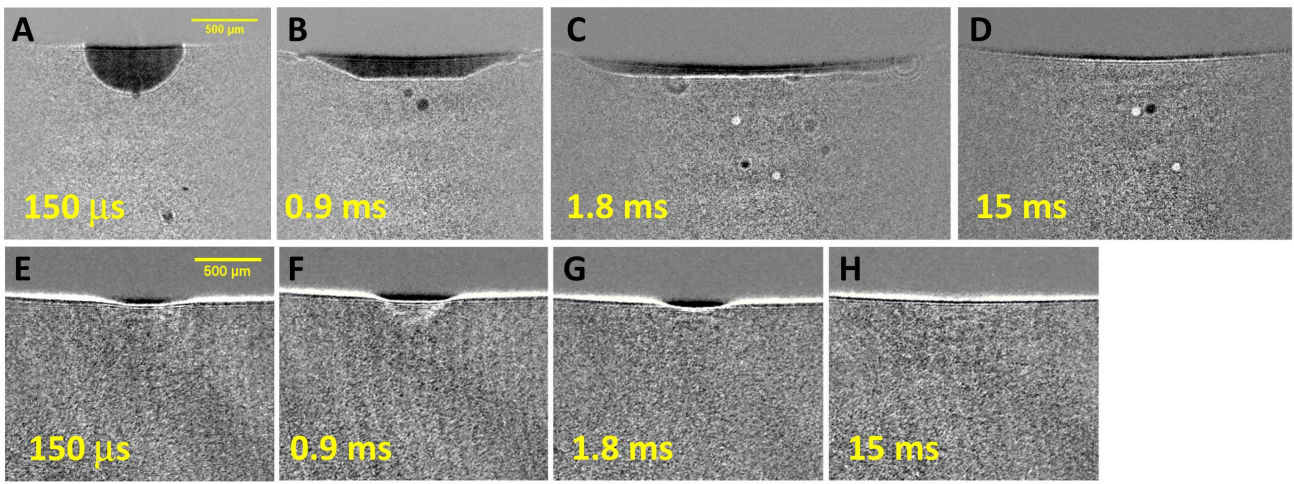

**Fig. 6.** THz-FEL-induced phenomena over a long time scale, with images of bubble formation and collapse induced by the THz pulse train shown in Fig. 2(**A**) with a frequency of 4 THz and an average micropulse energy of 18 μJ at delay times of (**A**) 150 μs, (**B**) 0.9 ms, (**C**) 1.8 ms, and (**D**) 15 ms. A series of images (**E**) to (**H**) shows the surface depression induced by the single THz pulse shown in Fig. 2(**B**) with a frequency of 5 THz. These images were taken with a time gate of 10 ns.

Because the increases in temperature and local pressure during photoacoustic wave propagation are very important features for chemical and biological reactions in practical applications, we must further investigate the photoacoustic wave propagation with thermodynamic and hydrodynamic simulations. Another significant feature in Fig. 4(b) is that there is no obvious threshold of the input THz energy for photoacoustic wave generation, which is in contrast to the conventional method triggered by plasma generation with a near-IR laser.[4] This is one of evidences that the THz-light-induced pressure wave is caused by strong linear absorption of THz light by water.

## Conclusion

In summary, we demonstrated generation and observation of a THz-light-induced photoacoustic wave. The initial process prior to photoacoustic wave generation was linear absorption of the THz light by water, in contrast with plasma generation at the air–water interface by multi-photon absorption of near- or mid-IR light.

This provides great advantages for the use of THz-induced photoacoustic waves compared to the conventional IR-induced methods. First, the linear absorption by water is a mild and non-destructive process, so the proposed method can be applied to biological tissue and fragile instruments. Next, loosely focused THz light can generate a photoacoustic wave with a plane wavefront because of effective energy transfer from the THz light to thermal energy, owing to strong linear absorption by water. The propagation of the plane wave is easily controlled with reflective and focusing elements. Third, the THz-light-induced photoacoustic wave requires a small incident radiant exposure of < 1 mJ/cm$^2$.

THz light at a frequency of 3–7 THz can directly excite the collective intermolecular vibration of water molecules, which may cause effective generation of a macroscopic pressure wave, in contrast with mid-IR light exciting the intramolecular vibration mode. As a sub picosecond pulse of several microjoules in the mid-IR to THz frequency region has recently been produced using a tabletop femtosecond laser system,[38,39] the photoacoustic wave generation system can potentially be moved from an FEL facility to a laser laboratory for medical and industrial applications. In addition, the influence of the excited molecular motion, the inter or intramolecular vibration of water, on the photoacoustic wave generation will be examined by this system. The numerical study is also important to evaluate and optimize the photoacoustic wave generation. We would like to find a model to describe the three-dimensional photoacoustic wave propagation induced by the THz light irradiation to water.

The large absorption coefficient of the THz light means that the penetration depth in water is considerably shorter than 1 mm. Therefore, the THz light can directly affect molecules or biological tissues only within a submillimeter range. In previous studies, THz-light-induced DNA damage to a human skin sample with a thickness of less than 0.1 mm has been examined and discussed.[40-43] THz-light-induced photoacoustic waves will potentially be able to probe and control chemical reactions and biological structures beyond the penetration depth.[21]

**References**
1   Abraham, E., Minoshima, K. & Matsumoto, H. Femtosecond laser-induced breakdown in water: time-resolved shadow imaging and two-color interferometric imaging. *Opt. Commun.* **176**, 441-452 (2000).
2   Schaffer, C. B., Nishimura, N., Glezer, E. N., Kim, A. M. T. & Mazur, E. Dynamics of femtosecond laser-induced breakdown in water from femtoseconds to microseconds. *Opt. Express* **10**, 196-203 (2002).
3   Vogel, A., Noack, J., Huttman, G. & Paltauf, G. Mechanisms of femtosecond laser nanosurgery of cells and tissues. *Appl. Phys. B* **81**, 1015-1047 (2005).
4   Strycker, B. D. *et al.* Femtosecond-laser-induced shockwaves in water generated at an air-water interface. *Opt. Express* **21**, 23772-23784 (2013).
5   Beard, P. Biomedical photoacoustic imaging. *Interface Focus* **1**, 602-631 (2011).
6   Wang, L. V. & Hu, S. Photoacoustic Tomography: In Vivo Imaging from Organelles to Organs. *Science* **335**, 1458-1462 (2012).
7   Taruttis, A. & Ntziachristos, V. Advances in real-time multispectral optoacoustic imaging and its applications. *Nature Photonics* **9**, 219-227 (2015).
8   Wang, L. V. & Yao, J. A practical guide to photoacoustic tomography in the life sciences. *Nat. Methods* **13**, 627-638 (2016).
9   Vogel, A. & Venugopalan, V. Mechanisms of pulsed laser ablation of biological tissues. *Chem. Rev.* **103**, 577-644 (2003).
10  Doukas, A. G. & Flotte, T. J. Physical characteristics and biological effects of laser-induced stress waves. *Ultrasound Med. Biol.* **22**, 151-164 (1996).
11  Kodama, T., Hamblin, M. R. & Doukas, A. G. Cytoplasmic molecular delivery with shock waves: importance of impulse. *Biophys. J.* **79**, 1821-1832 (2000).
12  Sato, Y. *et al.* Targeted DNA transfection into the mouse central nervous system using laser-induced stress waves. *J. Biomed. Opt* **10**, 060501 (2005).
13  Sato, S., Ando, T. & Obara, M. Optical fiber-based photomechanical gene transfer system for in vivo application. *Opt. Lett.* **36**, 4545-4547 (2011).

14  Shimada, T., Sato, S., Kawauchi, S., Ashida, H. & Terakawa, M. *Focusing of photomechanical waves with an optical lens for depth-targeted molecular delivery*. Vol. 8941 PWB (SPIE, 2014).
15  Bell, C. E. & Maccabee, B. S. Shock-wave generation in air and in water by $CO_2$ TEA laser radiation. *Appl. Opt.* **13**, 605-609 (1974).
16  Emmony, D. C., Geerken, B. M. & Straaijer, A. Interaction of 10.6 μm laser radiation with liquids. *Infrared Phys.* **16**, 87-92 (1976).
17  Ostrovskaya, G. V. & Shedova, E. N. Optical studies of shock and acoustic waves generated due to absorption of $CO_2$ laser radiation in water. *Izv. Akad. Nauk SSSR Ser. Fiz.* **61**, 1342-1352 (1997).
18  Zelsmann, H. R. Temperature dependence of the optical constants for liquid $H_2O$ and $D_2O$ in the far IR region. *J. Mol. Struct.* **350**, 95-114 (1995).
19  Heyden, M. *et al.* Dissecting the THz spectrum of liquid water from first principles via correlations in time and space. *Proc. Natl. Acad. Sci. USA* **107**, 12068-12073 (2010).
20  Moosavi-Nejad, S. F., Hosseini, S. H. R., Satoh, M. & Takayama, K. Shock wave induced cytoskeletal and morphological deformations in a human renal carcinoma cell line. *Cancer Sci.* **97**, 296-304 (2006).
21  Yamazaki, S. *et al.* Propagation of THz irradiation energy through aqueous layers: Demolition of actin filaments in living cells. *Sci. Rep.* **10**, 9008 (2020).
22  Hoshina, H. *et al.* Polymer morphological change induced by terahertz irradiation. *Sci. Rep.* **6**, 27180 (2016).
23  Nagai, M. *et al.* Luminescence induced by electrons outside zinc oxide nanoparticles driven by intense terahertz pulse trains. *New J. Phys.* **19**, 053017 (2017).
24  Makino, K. *et al.* Significant volume expansion as a precursor to ablation and micropattern formation in phase change material induced by intense terahertz pulses. *Sci. Rep.* **8**, 2914 (2018).
25  Kawase, K. *et al.* The high-power operation of a terahertz free-electron laser based on a normal conducting RF linac using beam conditioning. *Nucl. Instrum. Methods Phys. Res. Sect. A- Accel. Spectrom. Detect. Assoc. Equip.* **726**, 96-103 (2013).
26  Kawase, K. *et al.* Extremely high-intensity operation of a THz free-electron laser using an electron beam with a higher bunch charge. *Nucl. Instrum. Methods Phys. Res. Sect. A- Accel. Spectrom. Detect. Assoc. Equip.* **960**, 163582 (2020).
27  Nahata, A., Weling, A. S. & Heinz, T. F. A wideband coherent terahertz spectroscopy system using optical rectification and electro-optic sampling. *Appl. Phys. Lett.* **69**, 2321-2323 (1996).
28  Wu, Q. & Zhang, X. C. Design and characterization of traveling-wave electrooptic terahertz sensors. *IEEE J. Sel. Top. Quantum Electron.* **2**, 693-700 (1996).
29  Knippels, G. M. H. & van der Meer, A. F. G. FEL diagnostics and user control. *Nucl. Instrum. Methods Phys. Res. Sect. B- Beam Interact. Mater. Atoms* **144**, 32-39 (1998).
30  Wang, X., Nakajima, T., Zen, H., Kii, T. & Ohgaki, H. Damage threshold and focusability of mid-infrared free-electron laser pulses gated by a plasma mirror with nanosecond switching pulses. *Appl. Phys. Lett.* **103**, 191105 (2013).
31  Sze, S. M. & Ng, K. K. *Physics of Semiconductor Devices*.  (Wiley, 2006).
32  Yamashita, G. *et al.* Sensitive monitoring of photocarrier densities in the active layer of a photovoltaic device with time-resolved terahertz reflection spectroscopy. *Appl. Phys. Lett.* **110**, 071108 (2017).
33  Settles, G. S. *Schlieren and shadowgraph techniques: Visualizing phenomena in transparent media*. (Springer-Verlag, 2001).
34  Rogers, P. H. Weak‐shock solution for underwater explosive shock waves. *J. Acoust. Soc. Am.* **62**, 1412-1419 (1977).
35  Greenspan, M. & Tschiegg, C. E. Tables of the speed of sound in water. *J. Acoust. Soc. Am.* **31**, 75-76 (1959).
36  Vogel, A., Linz, N., Freidank, S. & Paltauf, G. Femtosecond-laser-induced nanocavitation in water: implications for optical breakdown threshold and cell surgery. *Phys. Rev. Lett.* **100**, 038102 (2008).
37  Zhang, D., Gökce, B. & Barcikowski, S. Laser synthesis and processing of colloids: fundamentals and applications. *Chem. Rev.* **117**, 3990-4103 (2017).
38  Vicario, C. *et al.* Generation of 0.9-mJ THz pulses in DSTMS pumped by a Cr:$Mg_2SiO_4$ laser. *Opt. Lett.* **39**, 6632-6635 (2014).

39  Vicario, C. *et al.* High efficiency THz generation in DSTMS, DAST and OH1 pumped by Cr:forsterite laser. *Opt. Express* **23**, 4573-4580 (2015).
40  Wilmink, G. J. & Grundt, J. E. Invited Review Article: Current State of Research on Biological Effects of Terahertz Radiation. *J. Infrared Millim. Terahertz Waves* **32**, 1074-1122 (2011).
41  Titova, L. V. *et al.* Intense THz pulses down-regulate genes associated with skin cancer and psoriasis: a new therapeutic avenue? *Sci. Rep.* **3**, 2363 (2013).
42  Titova, L. V. *et al.* Intense THz pulses cause H2AX phosphorylation and activate DNA damage response in human skin tissue. *Biomed. Opt. Express* **4**, 559-568 (2013).
43  Bogomazova, A. N. *et al.* No DNA damage response and negligible genome-wide transcriptional changes in human embryonic stem cells exposed to terahertz radiation. *Sci. Rep.* **5**, 7749 (2015).

**Acknowledgements**
This work was performed under the Cooperative Research Program of the Network Joint Research Center for Materials and Devices. We are grateful for the financial support of the QST President's Strategic Grant (Creative Research). We thank K. Furukawa for his technical support of the THz-FEL operation.

**Author contributions**
M.T., M.N., and H.H. conceived this study. M.T. conducted the experiments and analyzed the measured data. G.I. constructed and managed the THz-FEL facility in Osaka Univ. M.N. characterized the THz-FEL outputs. M.T. drafted the original manuscript, and all authors edited and reviewed the manuscript.

**Competing interests**
The authors declare no competing interests.

**Additional information**
**Supplementary** information accompanies this paper at …
**Correspondence** and requests for materials should be addressed to M.T.
**Publisher's note:** Springer nature remains neutral with regard to jurisdictional claim in published maps and institutional affiliations.